# Ballistic performance comparison of monolayer transition metal dichalcogenide $MX_2$ (M = Mo, W; X = S, Se, Te) MOSFETs

Jiwon Chang, Leonard F. Register and Sanjay K. Banerjee

Microelectronics Research Center, The University of Texas at Austin, Austin, TX 78758, USA

We study the transport properties of monolayer $MX_2$ (M = Mo, W; X = S, Se, Te) n- and p-channel metal-oxide-semiconductor field effect transistors (MOSFETs) using full-band ballistic non-equilibrium Green's function simulations with an atomistic tight-binding Hamiltonian with hopping potentials obtained from density functional theory. We discuss the subthreshold slope, drain-induced barrier lowering (DIBL), as well as gate-induced drain leakage (GIDL) for different monolayer $MX_2$ MOSFETs. We also report the possibility of negative differential resistance behavior in the output characteristics of nanoscale monolayer $MX_2$ MOSFETs.

## I.  INTRODUCTION

During the past few years, transition metal dichalcogenides (TMDs) have been intensively investigated for next generation nanoelectronic devices. TMDs with a chemical formula $MX_2$, where M is a transition metal atom and X is one of the chalcogens such as S, Se and Te, are constructed by stacking multiple X-M-X layers. One X-M-X layer (monolayer $MX_2$) consists of an M atom layer sandwiched between two X atom layers. The M-X bonding is strong covalent, but coupling between $MX_2$ monolayers is only by weak van der Waals forces.

Therefore, micromechanical exfoliation can be used to fabricate and isolate $MX_2$ monolayers [1,2]. Due to their near-two-dimensional (2-D) structure, TMD monolayers can provide a higher degree of electrostatic control than the conventional bulk materials, making them promising for low power switching and device scaling.

Both theoretical [3,4,5,6] and experimental [1] studies of various $MX_2$ (M = Mo, W; X = S, Se, Te) materials, found a strong dependency of their band structures on the number of X-M-X layers. Multilayer $MX_2$ has an indirect band gap while a direct band gap is observed in the monolayer $MX_2$. The direct band gap size of monolayer $MX_2$ (M = Mo, W; X = S, Se, Te) ranges from ~ 1.0 to ~ 2.0 eV which is suitable for CMOS logic device applications [1,3,4,5,6]. Recently, an n-channel metal-oxide-semiconductor field effect transistor (n-MOSFET) with a monolayer $MoS_2$ channel was reported with high mobility, high ON-OFF current ratio and ultralow standby power dissipation [7]. Integrated circuits based on bilayer [8] and monolayer $MoS_2$ [9] have shown the promise of $MoS_2$ for the digital logic applications as well. A high performance p-channel MOSFET (p-MOSFET) with a monolayer $WSe_2$ channel was also demonstrated experimentally [10]. Theoretical studies employing an effective mass Hamiltonian have been used to estimate the ballistic performance limits of monolayer $MX_2$ MOSFETs [11,12,13]. However, for a more accurate ballistic treatment, we use a full-band Hamiltonian of monolayer $MX_2$. In this work, we perform quantum transport simulations with atomic orbital-based tight-binding (TB) Hamiltonians obtained from density functional theory (DFT) to explore the performance of monolayer $MX_2$ MOSFETs. Previously, we had used this approach to study transport in monolayer $MoS_2$ n-channel MOSFETs [14]. In this work, we extend this approach to additional materials and to p-channel devices, as well as provide more detail discussion. Finally, while not simulated here, we will also discuss the effects of scattering.

## II. COMPUTATIONAL APPROACH

Figure 1(a) and 1(b) show the crystal structure of monolayer $MX_2$ consisting of two hexagonal planes of chalcogen (X) atoms and an intermediate hexagonal plane of transition metal (M) atoms. The hexagonal primitive unit cell and experimentally measured lattice parameters *a* and *c* are indicated in Figure 1(a) and 1(b). We employ the OPENMX code [15] based on the linear combination of numerical atomic-orbital basis sets and pseudo potentials to perform DFT calculations. We adopted the local density approximation (LDA) [16] for exchange-correlation energy functional and used a kinetic energy cutoff of 200 Ryd and *k*-mesh size of 7×7×1. Since it has been reported that DFT calculations employing the experimental lattice constants reproduce the band gap well [3,5,6,12,14,17], we constructed the monolayer $MX_2$ structure with experimental lattice parameters (Table 1) [18,19,20] in our calculations. The band structures of monolayer $MX_2$, along the high symmetry point K-Γ-M in the hexagonal Brillouin zone, are shown in Figure 2. For all monolayer $MX_2$ considered in this work, the direct band gap minimum occurs at K as predicted by previous studies [3,4,5,6]. The band gap size of each monolayer $MX_2$ agrees well with the other plane-wave based DFT calculations [3,4,5,6] and available experimental values [1]. Estimated effective masses around the conduction band (CB) minimum and the valence band (VB) maximum in the direction Γ–K for each monolayer $MX_2$ are summarized in Table 1. Effective masses of both electrons and holes tend to increase as the X atom becomes heavier for the same M atom. With the same X atom, both electron and hole effective masses of $WX_2$ are lighter than those of $MoX_2$. We note that in reality spin-orbit coupling breaks the spin degeneracy in the VB, such that the states within the two otherwise equivalent K valleys have opposite spins at the same energy, while within the same K valley there is an ~100 meV band-edge splitting between spins in MoS$_2$ [21]. However, we use spin degenerate band structures for our transport calculation similar to the previous studies [12,22], which should produce limited error for the purposes considered here.

The simulated device structure of monolayer MX$_2$ MOSFETs is illustrated in Figure 1(c). A mean free path of 15 to 22 nm is suggested in [11], based in part on the experimental work of [7], such that at least quasi-ballistic transport might be expected on this scale. We consider 15 nm channel length n- and p-MOSFETs, respectively. The undoped monolayer MX$_2$ rests on top of a 50 nm thick SiO$_2$ substrate and is gated through 3.5 nm thick HfO$_2$ (dielectric constant $\kappa = 25$) gate insulator. The source and drain are n-type and p-type for n-MOSFETs and p-MOSFETs, respectively, doped to a carrier concentration of $3.5\times10^{13}$ cm$^{-2}$. Effective doping strategies for these ultrathin 2-D materials are still under consideration. However, doping by vacancies or substitutional impurity atoms in the monolayer MoS$_2$ has been reported [23], and doping by surface adatoms or molecules have been demonstrated in TMD FETs as well as in graphene FETs [10,24,25,26]. The relative dielectric constant used for each monolayer MX$_2$ material considered in the work is listed in Table 1 [27].

For these transport studies, the monolayer MX$_2$ is divided into a series of rectangular unit cells marked with a green rectangular in Figure 1(b), with $x$ being the nominal transport direction. The TB Hamiltonian used for this purpose employs maximally localized Wannier functions (MLWFs) [28], with five centered about the M atom in each primitive unit cell and four centered about each of the two X atoms. The MLWFs and onsite through 3$^{rd}$ nearest neighbor hopping potentials are calculated directly from the DFT Kohn-Sham orbitals and potential using OPENMX. The corresponding thirteen TB energy bands within the monolayer TMD Brillouin zone, four CBs and nine VBs, more than cover all energies relevant to transport under all considered contact biases, and reproduce the original DFT band structure well over the entire range, as illustrated in Figure 2 over part of that range. We inject carrier wavefunctions into the device simulation region from the propagating plane-wave eigenmodes of the semi-infinite source and the drain leads, and use recursive scattering matrices to propagate the carrier wavefunctions through the device [29]. The incident plane-waves are resolved with uniform energy spacing $\Delta E < 2$ meV, and $N_y = 200$ uniformly separated values of $k_y$ were used to keep the associated energy spacing in the first conduction band small. The propagating wave-functions are normalized to an *incident* current density of $e(\Delta E/\pi\hbar)$ per incident mode per unit device width $\Delta W = N_y a_y$, assuming spin degeneracy within the energy band, consistent with Landauer-Büttiker theory. Total current is calculated by summing the transmitted current over all modes with a Fermi function weight. The transport calculations are solved together with Poisson's equation iteratively until self-consistency between the charge density and electrostatic potential is obtained. All simulations are performed at 300 K.

### III.  RESULTS AND DISCUSSION

Simulation results for monolayer $MX_2$ n-MOSFETs are presented in Figure 3 and 4. For all $MX_2$ monolayers considered here, good subthreshold behavior and limited short-channel effects are observed in the transfer characteristics, $I_{DS}$ vs. $V_{GS}-V_T$, of Figure 3. The subthreshold slope and drain-induced barrier lowering (DIBL) estimated for each monolayer $MX_2$ are shown in Table II. Among the monolayer $MoX_2$ ($MoS_2$, $MoSe_2$, $MoTe_2$) n-MOSFETs, $MoS_2$ shows the smallest subthreshold slope (~ 60 mV/dec) and DIBL (~ 10 mV/V). With a heavier X (Se, Te) atom form $MoSe_2$ and $MoTe_2$, both subthreshold slope and DIBL increase, but remain relatively small. This slight degradation can be explained by the larger dielectric constant of $MoSe_2$ and MoTe$_2$ compared to that of MoS$_2$. With a higher dielectric constant, the lateral electric field from the drain has more influence on the channel, leading to the increase of subthreshold slope and DIBL. However, with only a monolayer of TMD, the dielectric environment is dominated by the substrate and gate material, mitigating the detrimental effects of increasing TMD dielectric constant. The subthreshold slope and DIBL for two different monolayer WX$_2$ (WS$_2$, WSe$_2$) n-MOSFETs are alike due to the similarity in the dielectric constant and the effective mass, as seen in Table I. All WX$_2$ n-MOSFETs have a somewhat better subthreshold slope and DIBL compared with MoSe$_2$ and MoTe$_2$ n-MOSFETs consistent with the smaller dielectric constants. The subthreshold slope and DIBL of monolayer WS$_2$, WSe$_2$, and MoS$_2$ n-MOSFETs are found to be very close. For all monolayer MX$_2$ materials except monolayer MoTe$_2$, gate-induce drain leakage (GIDL), a potentially significant component of OFF-state leakage current in materials such as Si and common III-Vs, is not possible within the voltage ranges considered here due to their large band gaps. For the monolayer MoTe$_2$, however, the subthreshold currents starts to increase below $V_{GS}$–$V_T \approx -0.7$ V. Because of its relatively smaller band gap (~ 1.1 eV), as shown in Figure 2(c), there exists an overlap between CB and VB in the region between the channel and drain for low $V_{GS}$–$V_T$, which allows channel-to-drain band-to-band tunneling. (For the calculation of GIDL only, the potential profile obtained without GIDL—which is extremely limited such that it will have a negligible effect on charge density—is used, and the otherwise source lead injection boundary is moved to location of the top of the source-to-channel barrier to calculate the inter-band transport.)

The linear scale plots of $I_{DS}$ vs. $V_{GS}$–$V_T$ in Figure 3 exhibit significantly better transconductance at $V_{DS} = 0.5$ V for the WX$_2$ TMDs as compared to the MoX$_2$ TMDs in these ballistic simulations. Moreover, the MoX$_2$ TMDs show limited improvement in transconductance from $V_{DS} = 0.05$ V to $V_{DS} = 0.5$ V unlike the WX$_2$ TMDs. The reason for this difference becomes clear from Figure 4, where it is seen that the MoX$_2$ TMD devices exhibit substantial NDR, as previously discussed for MoS$_2$ [14], while the WX$_2$ TMD devices exhibit some but much less NDR. Until NDR onset, all devices show much the same transconductance, which is on the scale of 5 mA/µm/V at $V_{DS} = 0.2$ V

To illustrate the source of NDR, we consider zero transverse momentum, $k_y = 0$, for these illustrations for specificity, and $V_{DS}$ values of 0.0, 0.1, 0.2, 0.3, 0.4 and 0.5 V in a monolayer MoSe$_2$ n-MOSFET in Figure 5. Each $V_{DS}$ dependent subfigure (a)-(f) has two CB plots at the transverse mode $k_y = 0$ on the left-hand-side (LHS), one in the source (black color) and the other in the drain, except at $V_{DS} = 0.0$ V since the two CBs overlap. The right-hand-sides (RHSs) are the sum of the transmission probabilities for $k_y = 0$ as a function of energy per spin state. For this purpose only, to isolate the full band structure effects, we use a non-electrostatically-self-consistent piecewise linear potential approximation for the source, channel and drain to analyze transmission paths and calculate transmission probabilities for different drain biases. We focus on the right-propagating modes of source and drain at Fermi level, since the current flow in the ON-state primarily occurs near the Fermi level, which was taken as the zero energy reference, and the transmission probabilities though the device between them. In Figure 5, right propagating modes are indicated by arrows and numbers. From Figure 5(a), there are two such propagating modes in the source and drain, respectively. Since there is no potential variation along the device with $V_{DS} = 0$ V, there is perfect quantum mechanical transmission in each mode at the Fermi level, as also shown in Table III, and the sum over transmission probability is simply equal to the number of right-propagating modes in the source, which is two per spin state. With $V_{DS} = 0.1$ V, the CB in drain is shifted down by the applied bias, as shown in Figure 5(b).

There remain two corresponding outgoing propagating modes in the drain, although the momenta $k_x$ of modes in the drain is somewhat different from those in the source. From table III, however, the transmission probability of the source mode 1 (2) to the drain mode 1 (2) remains almost unity, with the modes in the drain being accessible to those in the source via simple semi-classically accessible trajectories. For $V_{DS}$ = 0.2 V, two additional outgoing states appear in the drain (modes 3 and 4), but there is little transmission probability to these states, as also shown in Table III, because carriers injected from source modes 1 and 2 cannot reach these latter drain modes semiclassically. The total transmission probability remains approximately two with drain modes 1 and 2 still reachable by simple semiclassical trajectories. The simulation result is similar for $V_{DS}$ = 0.3 V. However, by the time $V_{DS}$ reaches 0.4 V and continuing through $V_{DS}$ = 0.5 V, Figure 5(e) and 5(f), respectively, only one outgoing mode per spin is reachable via a semiclassical trajectory, drain mode 1 from source mode 1. As a result, the sum over transmission probabilities drops to approximately unity at the Fermi level. Moreover, even this remaining semiclassical trajectory is convoluted, requiring the electron to first accelerate toward the drain, then decelerate, turn around, accelerate back toward the source, decelerate again, and turn back around, before finally accelerating back to and out of the drain. Quantum mechanically, the overall right-propagating state nominally incorporates a superposition of two right-going waves of different wavelengths and one left-going wave of yet another wavelength over a significant portion of the channel, making for a less adiabatic transition between source and drain. While the transmission probabilities in Table III remain high here for this source-to-drain trajectory in MoSe$_2$, under similar conditions we have observed the transmission probability to drop well below 0.5 in MoS$_2$ and WeS$_2$. In addition, an overall velocity reduction in the channel associated with this convoluted trajectory can be expected to affect somewhat the self-consistent electrostatics considered in the simulations of Figure 4 but not in the calculations of Table III and Fig. 5. In the calculations of Figure 4(a), the drain current drops by *more than* 50% (~60% in the range 0.4 V ≤ $V_{GS}$−$V_T$ ≤ 0.8 V) from its peak value near $V_{DS}$ = 0.2 to its value at $V_{DS}$ = 0.5 despite the remaining semiclassical path (and any non-semiclassical/tunneling current, although the latter likely remains small based on Table III). For MoSe$_2$, both thresholds for eliminating the simple semiclassical paths occurs at $V_{DS}$ ~ 0.35 V, with the region of strong NDR in Figure 4(a) smeared out about this voltage by roughly ± 0.1 V. Such smearing is to be expected given the depth of the Fermi sea in the source, thermal smearing of the occupation probabilities in the source about the Fermi level, and the quantum nature of the transport calculations that generally makes for less abrupt transitions.

As seen in Figure 4(b), all MX$_2$ monolayer n-MOSFETs except WS$_2$ exhibit significant NDR behavior within the 0 ≤ $V_{DS}$ ≤ 0.5 V simulation range, but with different $V_{DS}$ ranges of NDR and different amounts of current reduction. In a similar way to above, we find that this $I_{DS}$ vs. $V_{DS}$ behavior in terms of both the region of NDR and associated amount of current reduction for each is entirely consistent with conduction band structure of each MX$_2$ monolayer, which is shown in Figure 6 for $k_y$ = 0. In each subfigure, (a)-(e), the source-to-drain energy level threshold = e$V_{DS}$ at which one of the semiclassical source-to-drain trajectories (again excluding spin degeneracy) at the source Fermi level becomes convoluted, as described above, is marked with a solid line (magenta online). The threshold at which the other semiclassical trajectory is simply eliminated is marked with a dashed line (light blue online). For MoSe$_2$ (Figure 6(b)), both thresholds and, thus, the center of strong NDR in the device simulations of Figure 4(a) and 4(b) occur near $V_{DS}$ = 0.35 V as discussed above. Similar behavior is obtained for MoTe$_2$ (Figure 6(c)) but with both thresholds and the resulting center of strong NDR occurring for $V_{DS}$ near 0.3 V.

For MoS$_2$, (Figure 6(a)) the two thresholds are slightly separated, however, with that for one semiclassical path becoming convoluted when $V_{DS}$ is slightly above 0.35 V and that for the other semiclassical trajectory vanishing when $V_{DS}$ is slightly above 0.4 V. The NDR onset $V_{DS}$ for MoS$_2$ MOSFETs in Figure 4 is correspondingly slightly greater than for MoSe$_2$ MOSFETs. Note that the overall trend is for these threshold voltages in MoX$_2$ monolayer MOSFETs to be reduced as the X atom becomes heavier, again consistent with the shift of the regions of NDR in Figure 4(b). In WX$_2$ monolayer MOSFETs, these two thresholds occur at significantly higher values of $V_{DS}$, $V_{DS} \approx 0.55$ V and 0.70 V, respectively for WS$_2$, and $V_{DS} \approx 0.40$ V and 0.65 V, respectively for WSe$_2$. Only the threshold for formation of a convoluted semiclassical path in the WSe$_2$ n-MOSFET occurs within the $V_{DS}$ range of the device simulations of Figure 4(b). As a result, of the WX$_2$ n-MOSFETs, only WSe$_2$ n-MOSFETs exhibits significant NDR in Figure 4(b), and with less associated current reduction than for any of the MoX$_2$ n-MOSFETs.

Figures 7 and 8 provide transfer and output characteristics, respectively, of monolayer MX$_2$ p-MOSFETs. As shown in Figure 7, subthreshold slopes close to the thermal limit (60 mV/dec) and very small DIBL are obtained in all monolayer MX$_2$ devices, again consistent with their 2-D nature. Table IV summarizes the subthreshold slope and DIBL for each MX$_2$ monolayer p-MOSFET. As for the n-MOSFETs, MoS$_2$ and MoSe$_2$ n-MOSFETs show the best subthreshold behavior characterized by the smallest subthreshold slope (~ 65 mV/dec) and DIBL (< 10 mV/V). For the MoTe$_2$ p-MOSFETs, the subthreshold characteristics are slightly degraded with an increased subthreshold slope (~ 70 mV/dec) and DIBL (~ 18 mV/V) due to the larger dielectric constant of MoTe$_2$ compared with those of MoS$_2$ and MoSe$_2$, but are still good. WSe$_2$ also exhibits a low subthreshold slope similar to that of MoS$_2$, but with a slightly increased DIBL (~ 10 mV/V). Unusual subthreshold behavior is observed for WS$_2$ in Figure 7(g) at very small currents, where the subthreshold slope falls below the nominally ideal value of 60 mV/dec in a small range around $V_{GS}-V_T = 0.5$ V at $V_{DS} = -0.5$ V. This behavior is a result of aligning the source-to-channel barrier edge with the small band gap within the valence band structure of $WS_2$ in the drain, which is in the vicinity of 1 eV below the reference Fermi level in Figure 9(d). As for the n-MOSFETs, only the monolayer $MoTe_2$ p-MOSFET shows GIDL. The subthreshold current begins to increase above $V_{GS}-V_T \approx 0.7$ V due to its relatively smaller band gap (~ 1.1 eV).

The $MoS_2$ and, more so, $MoTe_2$ p-MOSFETs transfer characteristics of Figure 7(b) and 7(f) show relatively limited improvement in ON-state transconductance from $V_{DS} = -0.05$ V to $-0.5$ V. The output characteristics of Figure 8(b) show that the reason for this limited improvement is substantial NDR in the $MoS_2$ and, again more so, $MoTe_2$ p-MOSFETs. This NDR can be related to the formation of convoluted semiclassical conduction paths between source and drain, as discussed previously for n-channel devices, near $V_{DS} = 0.5$ V and 0.3 V, respectively, and as shown in the super-cell VB structures of Figure 9(a) and 9(c) for $k_y = 0$. The energies $eV_{DS}$ at which these convoluted paths form for source injection at the Fermi level are marked with solid lines. For $MoSe_2$, $WS_2$ and $WSe_2$ (Figure 9(b), 9(d) and 9(e), respectively), the lines are more than 0.5 eV away from the Fermi level, and NDR is not observed for these materials in the simulation range of $-0.5$ V $\leq V_{DS} \leq 0$ V of Figure 8(b), accordingly.

We note that the region of NDR in $V_{DS}$ will depend on the location of the Fermi level within the band structure, which is used as the energy reference in Fig. 6 and Fig. 9 for example, and, therefore, the source and drain carrier concentrations. In these simulations, a degenerate carrier concentration of $3.5 \times 10^{13}$ cm$^{-2}$ is assumed in the source and drain. Higher or lower carrier concentrations, respectively, would reduce or increase the magnitude of the NDR onset $V_{DS}$ somewhat.

We also remind the readers that our calculations are performed in the ballistic limit of transport. Scattering would substantially affect the ON-state transconductance and, particularly, the discussed NDR behavior. Scattering will allow intra-band source-to-drain transport even when not otherwise possible (or at least improbable allowing for band-to-band tunneling) by dissipating energy in the channel, and perhaps by adding additional inter-band transport paths. Therefore, in addition to reducing ON-state current under lower $V_{DS}$, scattering should *increase* the current beyond the nominal NDR onset voltages and, thereby, reduce or eliminate the NDR. In this way, the differences in transconductance among materials at higher $V_{DS}$ also likely will be reduced substantially by scattering. However, conversely, even limited NDR could serve as a signature of quasi-ballistic transport in nanoscale TMD MOSFETs.

## IV. CONCLUSION

In summary, we used atomistic full-band quantum transport simulations with TB potentials obtained from DFT, to investigate the device performances of single gate monolayer $MX_2$ (M = Mo, W; X = S, Se, Te) MOSFETs. The 15 nm channel length devices exhibited good subthreshold slopes close to the ideal value of 60 mV/decade, as well as small DIBL due to the electrostatic control afforded by the 2-D nature of monolayer $MX_2$ materials. Moreover, the large band gaps of most monolayer $MX_2$ TMDs suppress GIDL. These full band ballistic quantum transport simulations also exhibit substantial NDR in the $I_{DS}$ vs. $V_{DS}$ output characteristics. The source of this NDR is consistent with variation in the nature and number of transport paths from the source to the drain as a function of the source-to-drain bias $V_{DS}$. However, scattering should moderate or perhaps eliminate the NDR in principle depending on scattering rates and device dimensions, so these simulations should be taken as only suggesting the possibility of NDR. Conversely, even limited NDR could serve as a signature of quasi-ballistic transport in nanoscale TMD MOSFETs. Also, experimentally significant or not, understanding this NDR will be important for interpreting future full-band ballistic simulations of TMD MOSFETs.

ACKNOWLEDGMENT

The authors acknowledge support from the Nanoelectronics Research Initiative supported Southwest Academy of Nanoelectronics (NRI-SWAN) center, Intel, and NSF NASCENT. We thank the Texas advanced computing center (TACC) for computational support.

**TABLE I**. Lattice constant, band gap, effective mass and dielectric constant of monolayer $MX_2$

| $MX_2$ | Lattice Constant | | Band Gap [eV] | Effective Mass | | Dielectric Constant |
| --- | --- | --- | --- | --- | --- | --- |
| | a [Å] | c [Å] | | electron ($m_e^*/m_e$) | hole ($m_h^*/m_e$) | |
| $MoS_2$ | 3.160 | 3.172 | 1.81 | 0.56 | 0.64 | 4.8 |
| $MoSe_2$ | 3.299 | 3.352 | 1.51 | 0.62 | 0.72 | 6.9 |
| $MoTe_2$ | 3.522 | 3.630 | 1.10 | 0.64 | 0.78 | 8.0 |
| $WS_2$ | 3.155 | 3.160 | 1.93 | 0.33 | 0.43 | 4.4 |
| $WSe_2$ | 3.286 | 3.376 | 1.62 | 0.35 | 0.46 | 4.5 |

**TABLE II**. Subthreshold slope and DIBL for monolayer $MX_2$ n-MOSFETs

| $MX_2$ | Subthreshold Slope [mV/dec] | DIBL [mV/V] |
| --- | --- | --- |
| $MoS_2$ | ~60 | ~10 |
| $MoSe_2$ | ~65 | ~15 |
| MoTe | ~70 | ~20 |
| $WS_2$ | ~60 | ~7 |
| $WSe_2$ | ~63 | ~10 |

**TABLE III.** Transmission probabilities between individual source and drain modes for different drain biases of $V_{DS}$ = 0.1, 0.2, 0.3, 0.4, and 0.5 in monolayer $MoSe_2$ n-MOSFETs.

| $V_{DS}$ [V] | | Probabilities | | | | |
| --- | --- | --- | --- | --- | --- | --- |
| **0.0** | Source Modes \ Drain Modes | 1 | 2 | | | |
| | 1 | 1.00000 | 0.00000 | | | |
| | 2 | 0.00000 | 1.00000 | | | |
| **0.1** | Source Mode \ Drain Modes | 1 | 2 | | | |
| | 1 | 0.99970 | 0.00000 | | | |
| | 2 | 0.00001 | 0.94617 | | | |
| **0.2** | Source Mode \ Drain Modes | 1 | 2 | 3 | 4 | |
| | 1 | 0.99901 | 0.00000 | 0.00001 | 0.00006 | |
| | 2 | 0.00003 | 0.96740 | 0.00273 | 0.00000 | |
| **0.3** | Source Mode \ Drain Modes | 1 | 2 | 3 | 4 | 5 |
| | 1 | 0.89504 | 0.00009 | 0.00264 | 0.01323 | 0.00526 |
| | 2 | 0.00004 | 0.96472 | 0.00178 | 0.00021 | 0.01254 |
| **0.4** | Source Mode \ Drain Modes | 1 | | | | |
| | 1 | 0.97366 | | | | |
| | 2 | 0.018171 | | | | |
| **0.5** | Source Mode \ Drain Modes | 1 | | | | |
| | 1 | 0.96542 | | | | |
| | 2 | 0.02242 | | | | |

**TABLE IV**. Subthreshold slope and DIBL for monolayer $MX_2$ p-MOSFETs

| $MX_2$ | Subthreshold Slope [mV/dec] | DIBL [mV/V] |
|---|---|---|
| $MoS_2$ | ~65 | <10 |
| $MoSe_2$ | ~65 | <10 |
| MoTe | ~70 | ~18 |
| $WS_2$ | ~65 | NA |
| $WSe_2$ | ~65 | ~10 |

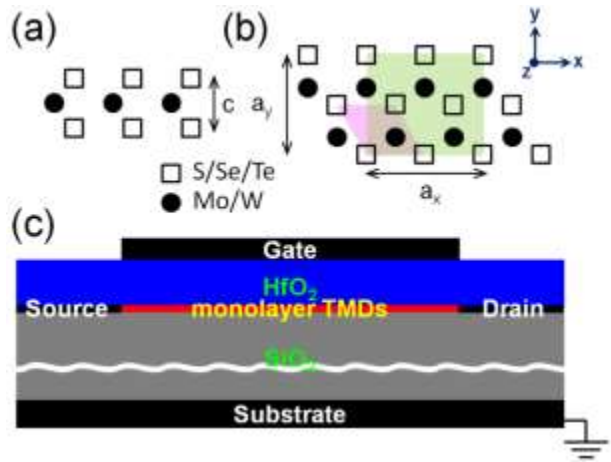

**Fig. 1**. (a) Side and (b) top views of monolayer TMDs. Hexagonal (magenta) and rectangular (green) unit cells are shown. Transport in the $x$ direction is considered. (c) Device structure of monolayer TMD MOSFETs. The nominal device parameters are as follows: $HfO_2$ ($\kappa = 25$) gate oxide thickness = 3.5 nm, channel length = 15 nm, n-type and p-type doping density of source and drain for n-MOSFETs and p-MOSFETs, respectively, of $3.5 \times 10^{13}$ cm$^{-2}$, and $SiO_2$ oxide thickness = 50 nm.

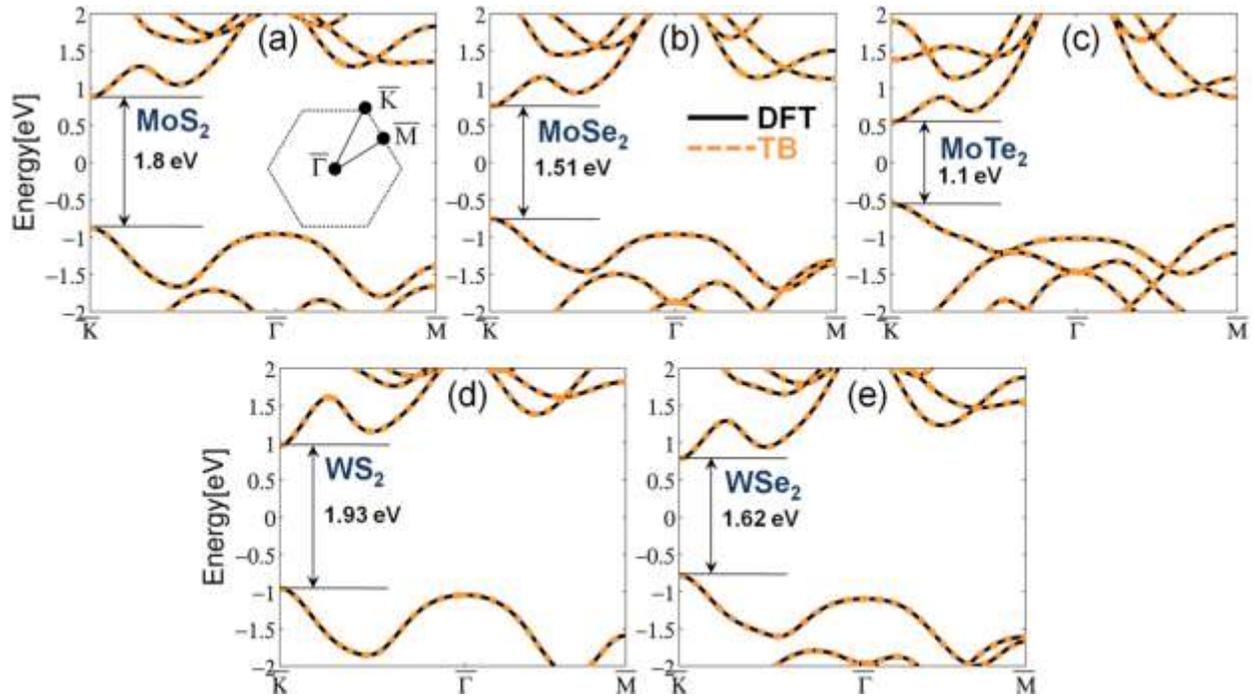

**Fig. 2**. Band structures of monolayer (a) $MoS_2$, (b) $MoSe_2$, (c) $MoTe_2$, (d) $WS_2$, and (e) $WSe_2$ calculated from DFT and via derived TB Hamiltonians, between the high symmetric points in the hexagonal Brillouin zone. A direct band gap at K is observed for each monolayer TMD.

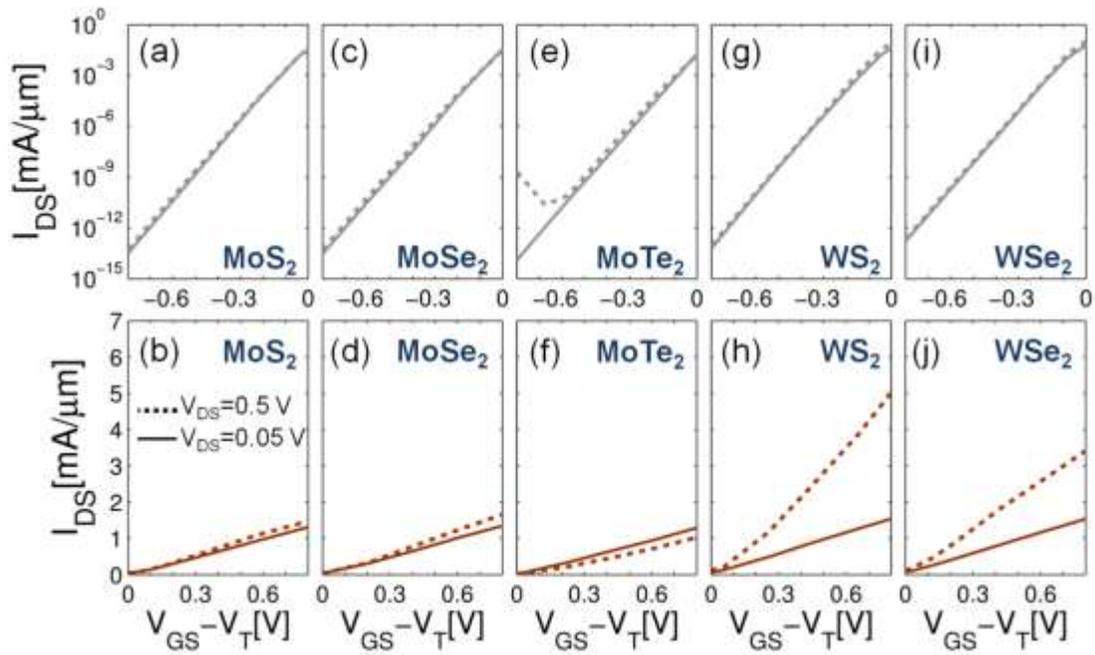

**Fig. 3**. $I_{DS}$ vs. $V_{GS}-V_T$ curves of 15 nm channel length monolayer $MX_2$ n-channel MOSFETs at $V_{DS}$ = 0.05 and 0.5 V, for (a), (b) $MoS_2$, (c), (d) $MoSe_2$, (e), (f) $MoTe_2$, (g), (h) $WS_2$, and (i), (j) $WSe_2$.

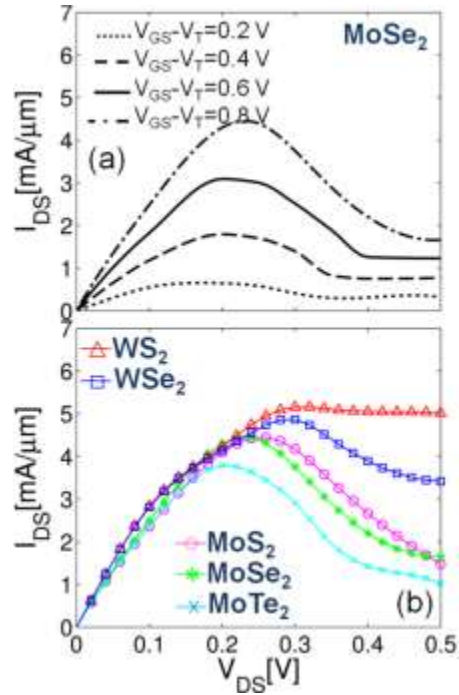

**Fig. 4**. (a) $I_{DS}$ vs. $V_{DS}$ curves of a 15 nm channel length monolayer $MoSe_2$ n-MOSFET at $V_{GS}-V_T$ = 0.2, 0.4, 0.6, and 0.8 V. (b) Comparison of $I_{DS}$ vs. $V_{DS}$ curves of 15 nm channel length monolayer $MX_2$ n-MOSFETs at $V_{GS}-V_T$ = 0.8 V.

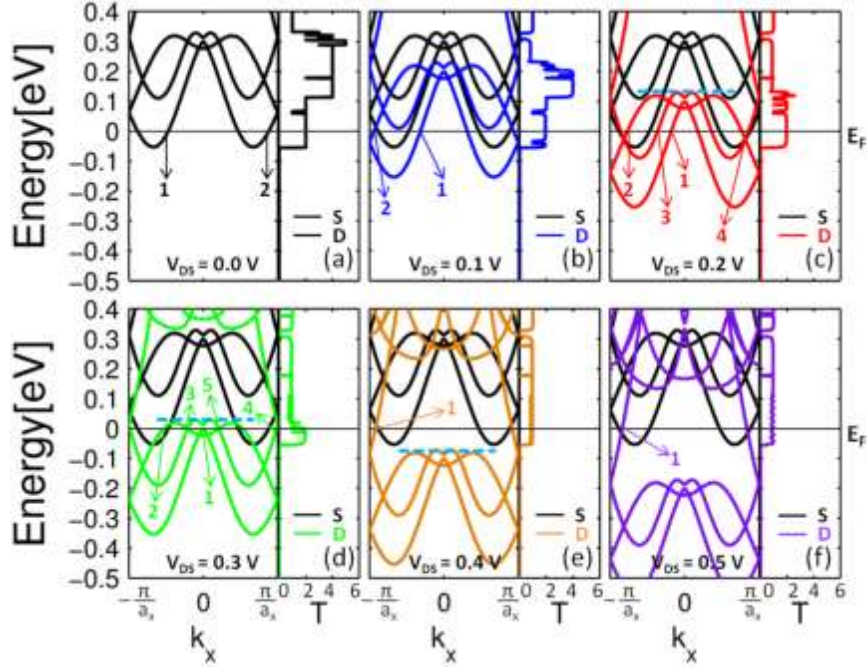

**Fig. 5**. CBs (LHS) in the source and drain and the sum over transmission probabilities (RHS) between the source and drain as a function of energy for a transverse crystal momentum $k_y = 0$ in monolayer MoSe$_2$ n-MOSFETs for drain biases of $V_{DS}$ = (a) 0.1, (b) 0.2, (c) 0.3, (d) 0.4, and (e) 0.5 V. Incoming modes in the source at the source Fermi level, which is used as an energy reference, are indicated by arrows and numbers in (a). The Fermi level position relative to the band edge is consistent with the assumed $3.5 \times 10^{13}$ cm$^{-2}$ source and drain doping density. Outgoing modes in the drain at *source* Fermi level are also indicated by arrows and numbers in (b), (c), (d), (e) and (f).

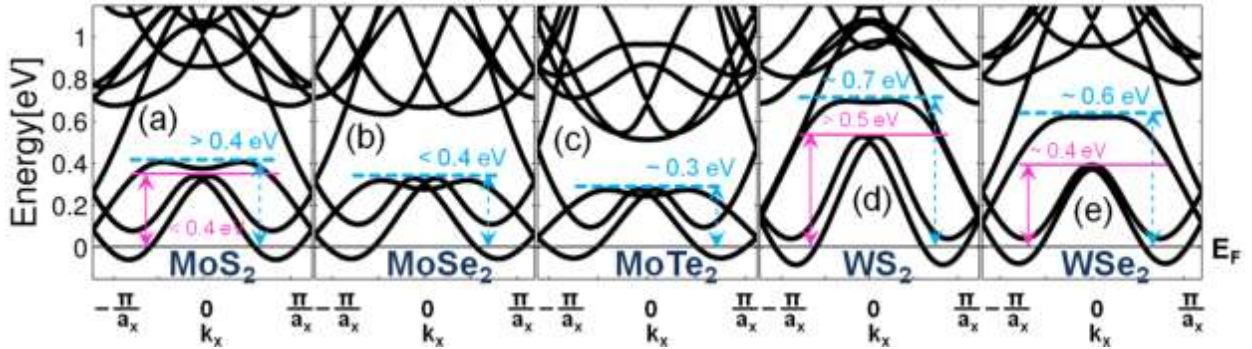

**Fig. 6**. Comparison of monolayer $MX_2$ CBs for a transverse mode $k_y = 0$. The energies of the solid (magenta on line) lines divide by unit of charge $e$ correspond to the source-to-drain voltages $V_{DS}$ at which one of the two classical trajectories per spin state becomes convoluted as described in the text; the dashed (light blue on line) lines correspond to the values of $V_{DS}$ at which the other semiclassical trajectory vanishes.

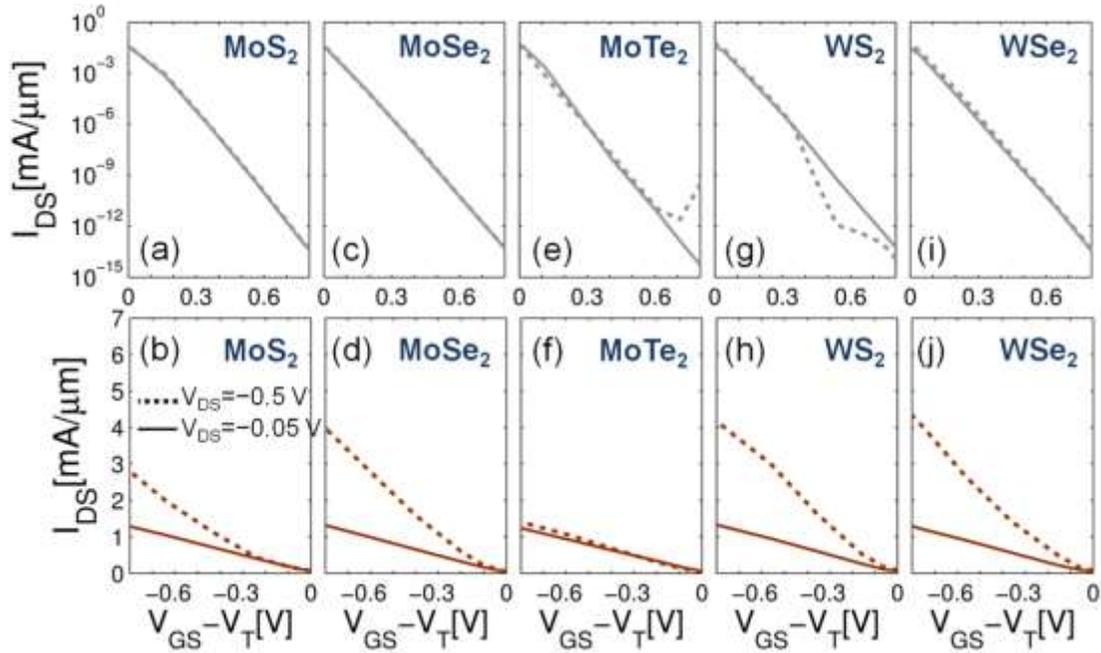

**Fig. 7**. $I_{DS}$ vs. $V_{GS}-V_T$ curves of 15 nm channel length monolayer $MX_2$ p-MOSFETs at $V_{DS}$ = −0.05 and −0.5 V, for (a), (b) $MoS_2$, (c), (d) $MoSe_2$, (e), (f) $MoTe_2$, (g), (h) $WS_2$, and (i), (j) $WSe_2$.

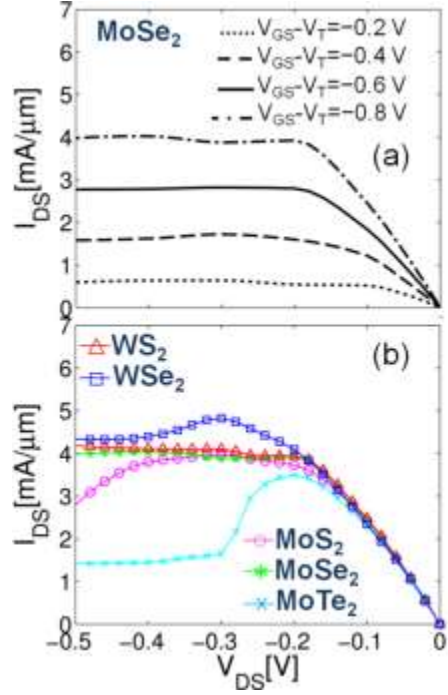

**Fig. 8**. (a) $I_{DS}$ vs. $V_{DS}$ curves of 15 nm channel length monolayer MoSe$_2$ at $V_{GS}-V_T = -0.2, -0.4,$ $-0.6,$ and $-0.8$ V. (b) Comparison of $I_{DS}$ vs. $V_{DS}$ curves of 15 nm channel length monolayer MX$_2$ at $V_{GS}-V_T = -0.8$.

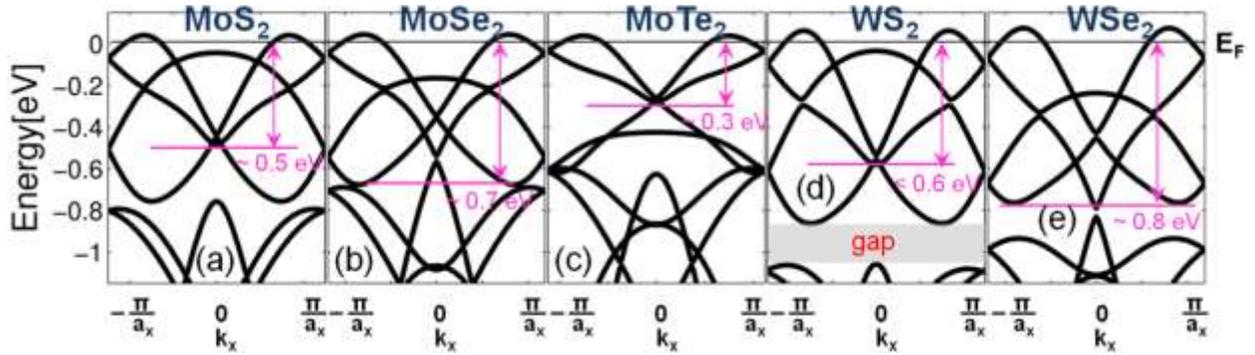

**Fig. 9**. Comparison of monolayer MX$_2$ VBs for a transverse mode $k_y = 0$. Fermi level is placed to dope each monolayer MX$_2$ to p-type with doping density $= 3.5 \times 10^{13}$ cm$^{-2}$.